\documentclass[journal=nalefd,manuscript=letter]{achemso}
\pdfoutput=1
\usepackage[T1]{fontenc} % Use modern font encodings

\usepackage{color}
\definecolor{bleufonce}{RGB}{33,57,138}
\definecolor{vertfonce}{RGB}{18,132,0}
\usepackage{amsmath}

\usepackage[numbers,sort&compress]{natbib}
\usepackage[colorlinks=true,linkcolor=bleufonce,citecolor=vertfonce]{hyperref}

\graphicspath{{Illustrations/}}

\author{Nicolas Chauvet}
\affiliation{Institut N\'eel, Univ. Grenoble Alpes, CNRS, Institut N\'eel}
\altaffiliation{Department of Information Physics and Computing, University of Tokyo}
\author{Maeliss Ethis de Corny}
\affiliation{Institut N\'eel, Univ. Grenoble Alpes, CNRS, Institut N\'eel}
\author{Mtthieu Jeannin}
\affiliation{Institut N\'eel, Univ. Grenoble Alpes, CNRS, Institut N\'eel}
\author{Guillaume Laurent}
\affiliation{Institut N\'eel, Univ. Grenoble Alpes, CNRS, Institut N\'eel}
\author{Serge Huant}
\affiliation{Institut N\'eel, Univ. Grenoble Alpes, CNRS, Institut N\'eel}
\author{Thierry Gacoin}
\affiliation{Laboratoire de Physique de la Mati\`ere condens\'ee, \'Ecole polytechnique}
\author{G\'eraldine Dantelle}
\affiliation{Institut N\'eel, Univ. Grenoble Alpes, CNRS, Institut N\'eel}
\author{Gilles Nogues}
\affiliation{Institut N\'eel, Univ. Grenoble Alpes, CNRS, Institut N\'eel}
\author{Guillaume Bachelier}
\affiliation{Institut N\'eel, Univ. Grenoble Alpes, CNRS, Institut N\'eel}
\email{guillaume.bachelier@neel.cnrs.fr}
\title{Hybrid KTP-plasmonic nanostructures for enhanced nonlinear optics at the nanoscale}


\begin{document}

%%%%%%%%%%%%%%%%%%%%%%%%%%%%%%%%%%%%%%%%%%%%%%%%%%%%%%%%%%%%%%%%%%%%%
%% The "tocentry" environment can be used to create an entry for the
%% graphical table of contents. It is given here as some journals
%% require that it is printed as part of the abstract page. It will
%% be automatically moved as appropriate.
%%%%%%%%%%%%%%%%%%%%%%%%%%%%%%%%%%%%%%%%%%%%%%%%%%%%%%%%%%%%%%%%%%%%%
%\begin{tocentry}
%
%Some journals require a graphical entry for the Table of Contents.
%This should be laid out ``print ready'' so that the sizing of the
%text is correct.
%
%Inside the \texttt{tocentry} environment, the font used is Helvetica
%8\,pt, as required by \emph{Journal of the American Chemical
%Society}.
%
%The surrounding frame is 9\,cm by 3.5\,cm, which is the maximum
%permitted for  \emph{Journal of the American Chemical Society}
%graphical table of content entries. The box will not resize if the
%content is too big: instead it will overflow the edge of the box.
%
%This box and the associated title will always be printed on a
%separate page at the end of the document.
%
%\end{tocentry}

%%%%%%%%%%%%%%%%%%%%%%%%%%%%%%%%%%%%%%%%%%%%%%%%%%%%%%%%%%%%%%%%%%%%%
%% The abstract environment will automatically gobble the contents
%% if an abstract is not used by the target journal.
%%%%%%%%%%%%%%%%%%%%%%%%%%%%%%%%%%%%%%%%%%%%%%%%%%%%%%%%%%%%%%%%%%%%%
\begin{abstract}

The search for miniaturized components for nonlinear optical processes needs a way to overcome the efficiency loss due to the effective size reduction of the active medium. We investigate here a combination of a nanosized nonlinear dielectric crystal and metallic nanoantennas that benefits from both the intrinsic nonlinear conversion efficiency of the nonlinear medium and the local-field enhancement of plasmonic resonances in metallic nanostructures. Careful comparison between experiments and numerical simulations reveals that the observed 10 to 1000 fold enhancement in Second Harmonic Generation intensity between isolated elements and their hybrid structure can be attributed unequivocally to the field enhancement effect of plasmonic resonances on the nonlinear crystal for gold - crystal structures, while the enhancement observed in aluminum-based hybrid structures is attributed to linear dielectric effect on the plasmonic antennas. %[ALUMINIUM]

%Combining the nonlinear efficiency of nanosized nonlinear dielectric material with the local-field enhancement of plasmonic resonance in metallic nanostructures is a promising way toward the miniaturization of optical components while limiting the efficiency drop due to size reduction. 

%Along with the current trend of optical component miniaturization, 

\end{abstract}

%%%%%%%%%%%%%%%%%%%%%%%%%%%%%%%%%%%%%%%%%%%%%%%%%%%%%%%%%%%%%%%%%%%%%
%% Start the main part of the manuscript here.
%%%%%%%%%%%%%%%%%%%%%%%%%%%%%%%%%%%%%%%%%%%%%%%%%%%%%%%%%%%%%%%%%%%%%
\section{Introduction}

%[ADD: SPDC?, QUANTITATIVE SIMULATIONS FOR DIRECT COMPARISON \\BETWEEN SYSTEMS, NO VARIABLE PARAMETER]

Nonlinear optical phenomena are at the core of many applications ranging from secure communication cryptography \cite{Gisin2007, Lo2014} to frequency conversion \cite{Eaton1991, Boyd2009} and quantum computing \cite{Kwiat1995, Horodecki2009, Barz2015}. Recent research works have focused on miniaturizing core elements, searching for ways to compensate the efficiency drop when the active medium reaches a sub-micron scale, such as waveguides \cite{Cowan2002} or plasmonic structures\cite{Rako1984}. This kind of object uses local field enhancement effects of plasmonic resonances in a nanoparticle to magnify nonlinear efficiency and far-field coupling within a dielectric material placed in its vicinity. By tailoring the shape and size of the plasmonic nanostructure, it is possible to obtain resonant behavior at several wavelengths simultaneously \cite{EthisDeCorny2016}, theoretically improving even more the conversion efficiency of frequency conversion processes (sum-frequency generation or down-conversion alike).

Several experiments have been reported in the last decades, which apply this principle to so-called hybrid structures, associating the intrinsic nonlinear efficiency of dielectric materials with the resonant behavior of plasmonic structures. Regarding second-order nonlinear processes, attempts include core-shell particles\cite{Pu2010, Ren2014}, dielectric nanowires and plasmonic particle arrays\cite{Grinblat2014a}, or controlled successive deposition of colloidal particles into an array of nano-holes on a substrate\cite{Timpu2017}. However, although these structures have indeed shown an improved nonlinear efficiency with respect to an isolated plasmonic or a dielectric nano-object, doubts have been expressed on the actual origin of this enhancement, with some results indicating that nonlinear effects within plasmonic structures alone actually overcome those of dielectric materials\cite{Linnenbank2016,Hentschel2016}.

In this work we experimentally identify the origin of Second Harmonic Generation (SHG) efficiency enhancement within individual hybrid nanostructures by carefully comparing the response of each sub-component to that of the hybrid systems, for two different plasmonic materials, gold or aluminum. Unique fabrication method as well as versatile, computer-controlled experimental setup have been developed to get as controlled and reproducible experimental conditions as possible. Quantitative numerical simulations based on Finite Element Method (FEM) enable us to compare the nonlinear contribution of each component separately without any adjusting parameter, thus helping to interpret the experimental data and to pinpoint the exact role of each nano-object on the nonlinear behavior of the whole hybrid system.

%Starting with nonlinear optics and second harmonic generation, this introduction will start from current challenges in component miniaturization to show the interest of plasmonic-based nonlinear hybrid structures.

%After a brief initial talk about plasmonics and its use, a litterature review dedicated to similar kinds of hybrid nanostructures will show the weaknesses in previous reports' conclusions and draw the outline of the current paper's strategy: under rigorously characterized conditions, show that a plasmonic structure placed in the vicinity of a nonlinear material can enhance the latter's conversion efficiency, and attribute unequivocally this enhancement from the (linear) plasmonic properties of the structure - and not its nonlinear ones.

\section{Results and discussion}

%[commencer par antennes, avec REF, etude numerique, double resonance, technique lithographie, electronique...]

%[\textit{REMANIER: antennes plasmoniques, effets NL, etudes/applications(?), contributions, RS, determination experimentale, simus quantitatives de chacune -> determiner laquelle domine, OK pour double reso alu, permet de predire quantitativement -> compatible avec autres contributions NL}]

%The hybrid structures include plasmonic nano-antennas (Al, Au) which are able to produce SHG by themselves, the latter being characterized by Rudnick and Stern parameters reported in the literature for bulk gold \cite{Bachelier2010} or aluminum \cite{Murphy1989} films. These parameters enable quantitative prediction of the SHG response from a single nanostructure without needing any adjusting parameter using finite element method simulations. We have previously reported on the nonlinear SHG response of individual plasmonic nanoantennas made by electronic lithography. In particular, we have demonstrated a double resonant regime in single Al nanoantennas, that enhances the SHG nonlinear efficiency when both resonances match the wavelengths involved in the process. Furthermore agreement with the numerical simulation is excellent and allows discriminating among the different processes at the origin of the SHG signal \cite{EthisDeCorny2016}.

Metallic films and nanoparticles have been known to have a second order nonlinear response since the discovery of laser light \cite{Bloembergen1962, Jha1965}. Because metals have a centrosymmetric structure, classical bulk description of second order nonlinear effects cannot be applied, and other explanations have to be found for the origin of their nonlinearity. Two main contributions to the nonlinear polarisation have been identified\cite{Bachelier2010}, one being the breaking of centrosymmetry at the surface of metals, which gives rise to a surface nonlinear polarisation $\vec{P}_{surf,\perp}(\vec{r},2\omega)$ under an electromagnetic excitation $\vec{E}(\vec{r},\omega)$ of the form:

\begin{equation}
\vec{P}_{surf,\perp}(\vec{r},2\omega)=\chi_{\perp\perp\perp}E_\perp(\vec{r},\omega)E_\perp(\vec{r},\omega)\vec{n}
\end{equation}
where $\vec{n}$ is the unitary vector normal to the surface at point $\vec{r}$ and $E_\perp$ is the projection of $\vec{E}$ along this direction. The second one can be found by using a hydrodynamic model for the equation of motion of a conduction electron in the metal \cite{Sipe1980}, which leads to another term of nonlinear polarisation within the bulk of the metal:

\begin{equation}
\vec{P}_{bulk}(\vec{r},2\omega)=\gamma_{bulk}\nabla.[\vec{E}(\vec{r},\omega).\vec{E}(\vec{r},\omega)]
\end{equation}
Here, $\vec{E}$ and $\vec{P}_{bulk}$ are evaluated inside the metal, while $\vec{P}_{surf,\perp}$ is evaluated just outside the metal. Following the treatment of Sipe et al. \cite{Sipe1980}, the nonlinear coefficients $\chi_{\perp\perp\perp}$ and $\gamma_{bulk}$ are defined in our model and simulations by \cite{Bachelier2010}:
\begin{align}
\chi_{\perp\perp\perp}&=-\dfrac{a}{4}[\varepsilon_r(\omega)-1]\dfrac{e\varepsilon_0}{m\omega^2}\label{eq:surf}\\
\gamma_{bulk}&=-\dfrac{d}{8}[\varepsilon_r(\omega)-1]\dfrac{e\varepsilon_0}{m\omega^2}
\label{eq:bulk}
\end{align}
where $\varepsilon_0$ is the vacuum permittivity, $\varepsilon_r$ the relative permittivity of the metal, $m$ and $e$ the mass and charge of the electron respectively, and $a$ and $d$ the adimensional Rudnick and Stern parameters \cite{Rudnick1971}. We use values of these parameters taken from the literature \cite{Murphy1989,Wang2009} in order to get quantitative predictions in terms of photons/s that can then be directly compared with experiments without any adjustable free parameter. This enables us to discriminate the effect of different contributions, in order to verify which one dominates in a given metallic nanostructure and to predict (double) resonant behavior \cite{EthisDeCorny2016}. As such, surface contribution (coefficient $a$) has been shown to dominate in aluminum structures, whereas bulk contribution (coefficient $d$) dominates in gold lithographed films \cite{Wang2009}. In this study, only the dominant nonlinear contribution is considered for each metal (Al, Au), which can then be compared directly with other nonlinear sources.

\begin{figure}[!htbp]
  \centering\includegraphics[scale=.7]{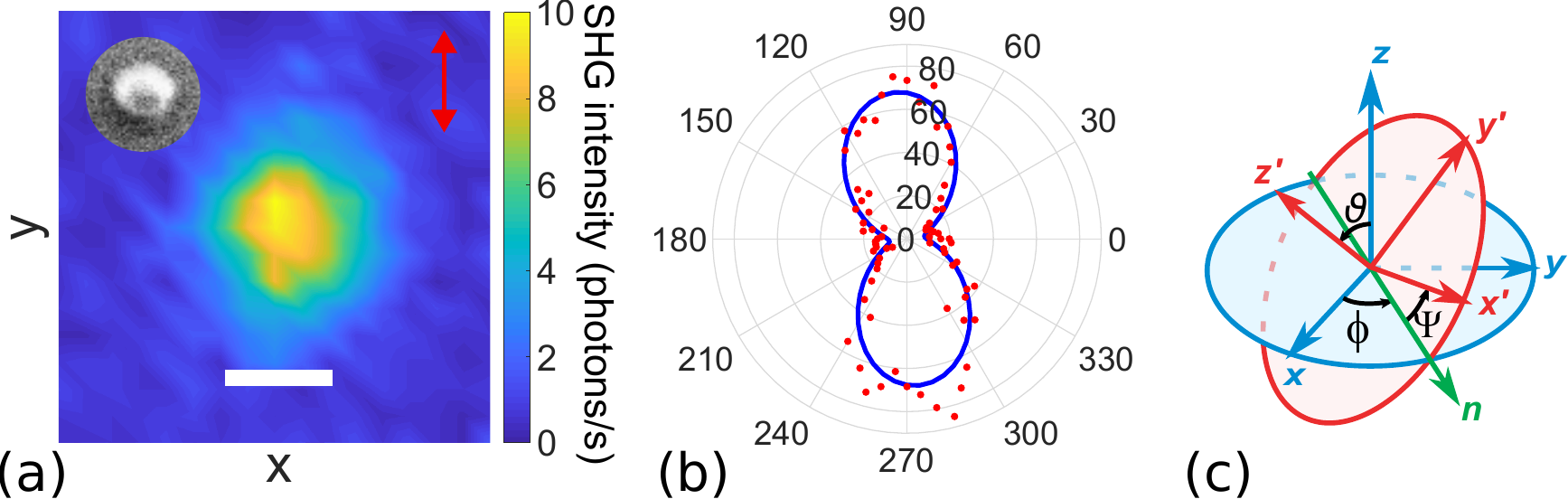}
  \caption{Single KTP nanocrystal analysis in SHG following previous works \cite{Brasselet2004, Mayer2013}. (a) SHG excitation cartography of an individual crystal whose SEM image is shown in insert (scale bar: 100 nm for both images). (b) SHG response in photons/s (with 100 $\mu$W of average excitation power) with respect to the linear excitation polarization angle from $x$ axis, which can be used to find (c) its 3 dimensional crystalline orientation through its Euler angles; in the displayed case, $(\theta,\phi,\psi)$=(50$^{\circ}$,130$^{\circ}$,20$^{\circ}$) in the laboratory reference frame.}
  \label{fig:crystal}
\end{figure}

In the structures investigated here, a nanocrystal made of nonlinear material is inserted between two identical antennas, either in gold or aluminum. The crystal itself has to meet several criteria to be used in our experiments. For this study, we have chosen potassium titanyl-phosphate (KTP) nano-sized ($\sim$80$\pm$20 nm) crystals \cite{Biswas2007, Mayer2013}, mainly because of their transparency range including the near UV, their high nonlinear efficiency and robustness against high temperature, hygrometry, optical power and electron beams \cite{Zumsteg1976, Eaton1991}. Under an incoming electromagnetic excitation $\vec{E}$ at frequency $\omega$, a non-centrosymmetric material shows a second order nonlinear polarization of the form:
\begin{equation}
P^{(2)}_{i}(2\omega;\omega,\omega,\vec{r})=\varepsilon_0\sum_{j,k}\chi^{(2)}_{ijk}(2\omega;\omega,\omega)E_j(\omega,\vec{r})E_k(\omega,\vec{r})
\end{equation}
where $i,j,k$ represent the polarization directions following the crystallographic axes ($x'$,$y'$,$z'$). Like most nonlinear materials, KTP has an anisotropic nonlinear response that is best stimulated if the incoming light is polarized along a specific crystallographic direction, in this case the $z'$ axis. In order to make a hybrid structure optimized for SHG, it is then necessary to determine the 3-dimensional crystallographic orientation to obtain the highest possible nonlinear response. Other studies have been made on individual KTP nanocrystals in SHG, including a technique based on far-field polarization measurement \cite{Mayer2013} to determine the orientation of the nonlinear susceptibility tensor of any crystal in the laboratory referential frame through its Euler angles. Figure \ref{fig:crystal} represents the typical different steps carried out on each nanocrystal with a pulsed 850 nm laser excitation beam focused through an immersion oil objective onto the sample (see Experimental section for details). The excitation cartography in Figure \ref{fig:crystal}(a) is obtained by scanning the sample with respect to the optical axis of the laser and collecting the emitted nonlinear intensity in the far-field through the same objective. The high numerical aperture (NA=1.3) of the objective yields a resolution of about 330 nm in excitation and half of it at the harmonic wavelength, which guarantees observations at the single particle level.

Following the method developped in \cite{Brasselet2004} and already applied to KTP \cite{Mayer2013}, we plot in Figure \ref{fig:crystal}(b) an analysis of the SHG response as a function of the excitation polarization. A fit to the data based on analytic simulations taking into account the strong focusing gives access to each crystal Euler angles as defined in (c). This information can then be integrated into numerical simulations to find the orientation of plasmonic antennas that optimizes the nonlinear efficiency of the hybrid system. 

In accordance with our objective to compare the nonlinear response of each component of the hybrid structure taken individually, we adopted a step-by-step approach: after performing SHG polarization analysis on 122 individual crystals, optimized plasmonic antennas are fabricated in the vicinity of each of them with the required angle and geometry by electron beam lithography (see Experimental section). Scanning Electronic Microscopy (SEM) realignment technique enables us to get an alignment precision of the order of 20-30 nm on the absolute position of the antennas, thus allowing for embedding a single nanocrystal in the gap between two Al or Au antennas.

\begin{figure}[!htbp]
  \centering\includegraphics[scale=.7]{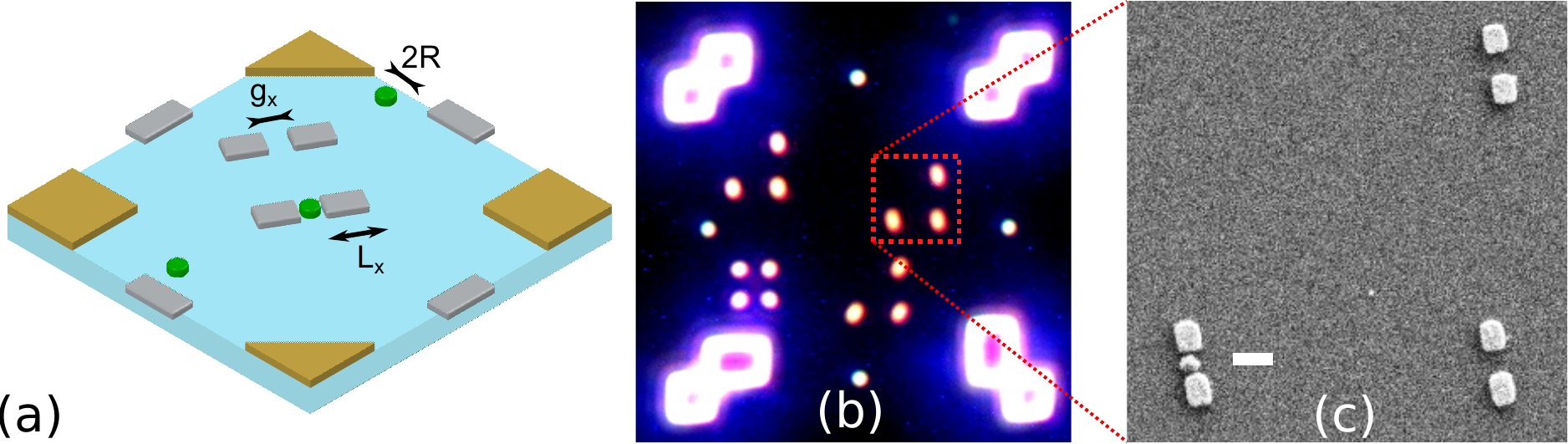}
  \caption{(a) Scheme of the sample design with relevant parameters referenced in the text. Gold reference marks are in brown, aluminum antennas in grey and KTP nanocrystals in green. (b) Dark field optical image of a sample after e-beam lithography with visible reference structures for nanofabrication quality check, with (c) a detailed SEM view of an hybrid gold-KTP structure (bottom-left corner). Scale bar: 200 nm.}
  \label{fig:fabrication}
\end{figure}

The design of the sample is represented on the scheme of figure \ref{fig:fabrication}(a). Au alignment marks were fabricated along a 10-$\mu$m stepped grid during a first lithography step to serve as reference points for SEM realignment, after which KTP nanocrystals were spin-coated before SHG study and SEM imaging of more than 100 individual nanocrystals to get their coordinates in the sample frame. After this thorough step, another electronic lithography step was used to fabricate the plasmonic antennas with desired parameters (angle, size, gap, position) around the crystal. During this step, identical plasmonic antennas were also realized nearby to get a nonlinear intensity reference without the crystal in the gap. Figure \ref{fig:fabrication}(b) shows dark-field optical image of the obtained sample, where Au antennas have been lithographed between alignment marks, while figure \ref{fig:fabrication}(c) shows a higher resolution SEM image. The hybrid structure is clearly seen at the bottom left of the SEM image, while replicas with identical fabrication parameters are located on the right without crystal.

%[UPDATE WITH NEW VALUES FOR AL] The final step consists in optimizing the relevant parameters of the plasmonic antennas to increase the nonlinear efficiency of the whole hybrid structure, by using the FEM numeric tools described before. Figure 3 shows the expected SHG intensity enhancement between a single 46 nm-wide, 40 nm-thick cylindrical KTP crystal, and a hybrid structure with the same crystal plus two symmetric plasmonic antennas made of (a) aluminum or (b) gold, as a function of the main axis length $L_x$ and the antenna gap $g_x$ along this same axis, while their width and thickness are fixed at 100 and 35 nm, respectively. Maximum enhancement factors are predicted at 867-fold for aluminum and 533-fold for gold, with different antenna lengths (200 and 150 nm respectively) but with the same, minimal antenna gap width. This confirms that the near-field coupling between antennas and crystal is essential to explain the predicted nonlinear enhancement effect. The SEM images shown in figure 3(c) indicates that the simulated geometry is compatible with actual observed configurations, although some of the aluminum antennas have a dilated shape like ii. because of a defocused electron beam during the lithography.

\begin{figure}[!htbp]
  \centering\includegraphics[scale=.7]{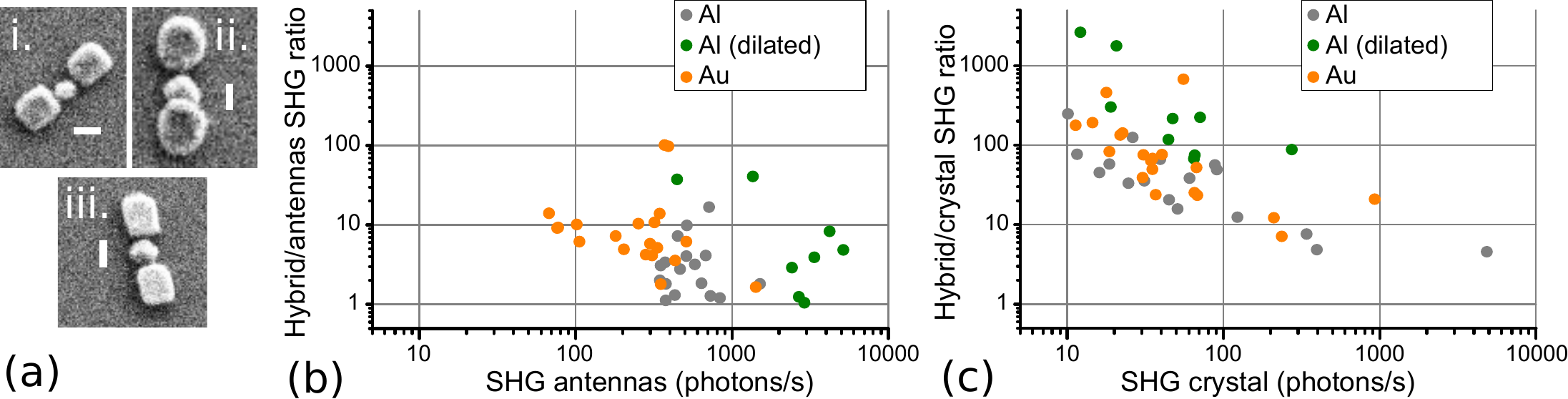}
  \caption{(a) SEM images of i. Al-KTP hybrid, ii. Al-KTP hybrid made with unfocused e-beam, iii. Au-KTP hybrid. Scale bar: 100 nm. (b) Ratio between the SHG of the hybrid structures and of the bare antennas with respect to the latter. (c) Statistics of the SHG enhancement ratio between hybrid and its bare crystal for different kinds of structures with respect to the bare crystal intensity.}
  \label{fig:stats}
\end{figure}

The antenna parameters are determined from SEM images of the hybrid structures. Three sets of systems have been realized as shown in figure \ref{fig:stats}(a): \textbf{i.} Al-KTP structures in close agreement with the targeted geometry, \textbf{ii.} Al-KTP with elliptical antennas due to an unfocussed e-beam and \textbf{iii.} Au-KTP structures also close to the targeted geometry. An average 110 nm width and 145 nm length have been measured for the antennas of the sets \textbf{i.} and \textbf{iii.}, while corresponding total gap distances are of the order of the diameter of the crystal plus 20 nm. In the case of set \textbf{ii.}, the aluminum antennas have more elliptical shapes closer to 140 nm width by 190 nm length. Their gap distance is closer to the crystal diameter, sometimes even with an overlap with the crystal.

For each hybrid structure, we compare the maximum SHG intensity experimentally measured on the single crystal (before the electronic lithography), the bare plasmonic antennas and the hybrid itself under identical experimental conditions (configuration, laser wavelength, power, and input polarization). All 80 studied hybrid systems have higher SHG signals than individual components. In particular, Figure \ref{fig:stats}(b) shows the ratio between the SHG intensities from the hybrid structures and the corresponding bare antennas as a function of maximum intensity from the antenna system, for hybrid antenna sets \textbf{i.} in grey (regular Al antennas), \textbf{ii.} in green (elongated Al antennas) and \textbf{iii.} in orange (regular Au antennas). SHG signal collected from the bare antennas is higher for aluminum antennas compared to gold, especially for set \textbf{ii.} In addition, the average SHG enhancement factor between bare plasmonic antennas and hybrid structures is about 11 for set \textbf{i.}, 4 for set \textbf{ii.} and 16 for set \textbf{iii.}. On the other hand, Figure \ref{fig:stats}(c) presents the ratio between the SHG intensities from the hybrid structure and the corresponding nanocrystal as a function of the nanocrystal SHG intensity. We observe that the intensity distribution of bare crystals covers nearly 3 orders of magnitude (due to size dispersion) while it is narrower for each set of bare antennas (for which the dimension is controlled by lithography). Enhancement factors reach several hundreds, even in the case of Au antennas which do not have plasmonic resonance at SHG wavelength. The best enhancement is clearly obtained for set \textbf{ii.}, indicating that the shape and gap size play an important role on the hybrid structure efficiency. Note that enhancement factors are by construction larger for smaller nanocrystal sizes (and therefore smaller intensities).

\begin{figure}
 \centering\includegraphics[scale=.9]{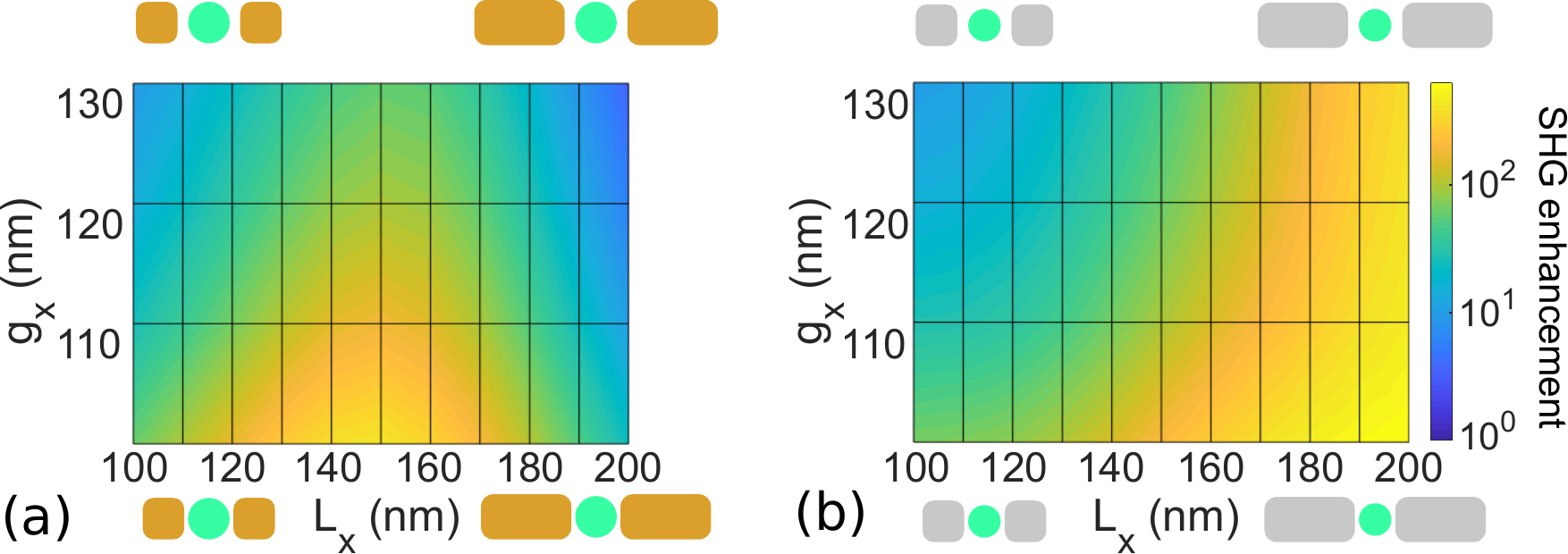}
 \caption{\small Numerical simulations of the predicted SHG enhancement in logarithmic scale with respect to the bare crystal for (a) Au-KTP or (b) Al-KTP hybrid structure, as a function of the length $L_x$ of both antennas and their edge-to-edge distance $g_x$ along their main axis. Width is fixed at 100 nm, thickness at 35 nm.}
 \label{fig:optimisation}
\end{figure}

To understand the influence of the hybrid system parameters on its nonlinear efficiency, quantitative FEM numeric simulations\cite{Bachelier2008,Bachelier2010,EthisDeCorny2016} were used and compared to experimental data. Figures \ref{fig:optimisation} (a) and (b) show the simulated SHG intensity enhancement between a single 46 nm-wide, 40 nm-thick cylindrical KTP crystal, and the corresponding hybrid structure of (a) gold and (b) aluminum. Antenna width and thickness are fixed to 100 and 35 nm respectively while their length $L_x$ and gap $g_x$ are varied. The simulations use the crystallographic orientation of an observed crystal as determined by the method described before with $(\theta,\phi,\psi)$=(130$^{\circ}$,0$^{\circ}$,20$^{\circ}$), assuming that the antenna main axis is aligned with the projection of the crystallographic $z$ axis in the $xy$ plane of the laboratory frame (or equivalently, $\phi=0^{\circ}$). Maximum enhancement factors with respect to the bare crystal are predicted at 640 for Al and 530 for Au, which are within the range of experimental values of Figure \ref{fig:stats}(c) for the sets \textbf{ii.} (elongated Al) and \textbf{iii.} (Au). They occur for different antenna lengths (200 and 150 nm respectively) but always with a minimal antenna gap. This confirms that the near-field coupling between antennas and crystal is essential to explain the predicted nonlinear enhancement effect. Besides, this also explains why set \textbf{i.} (regular Al antennas) shows inferior values to set \textbf{ii.}, since the elongated antennas are both closer to the crystals (or even in contact) and with lengths closer to 200 nm. Further simulations have confirmed that 200 nm-long antennas indeed show the best enhancement factor for aluminum.
%[DISCUSSION A FINIR EN FONCTION DE LA FIGURE]

\begin{figure}[!htbp]
  \centering\includegraphics[width=\textwidth]{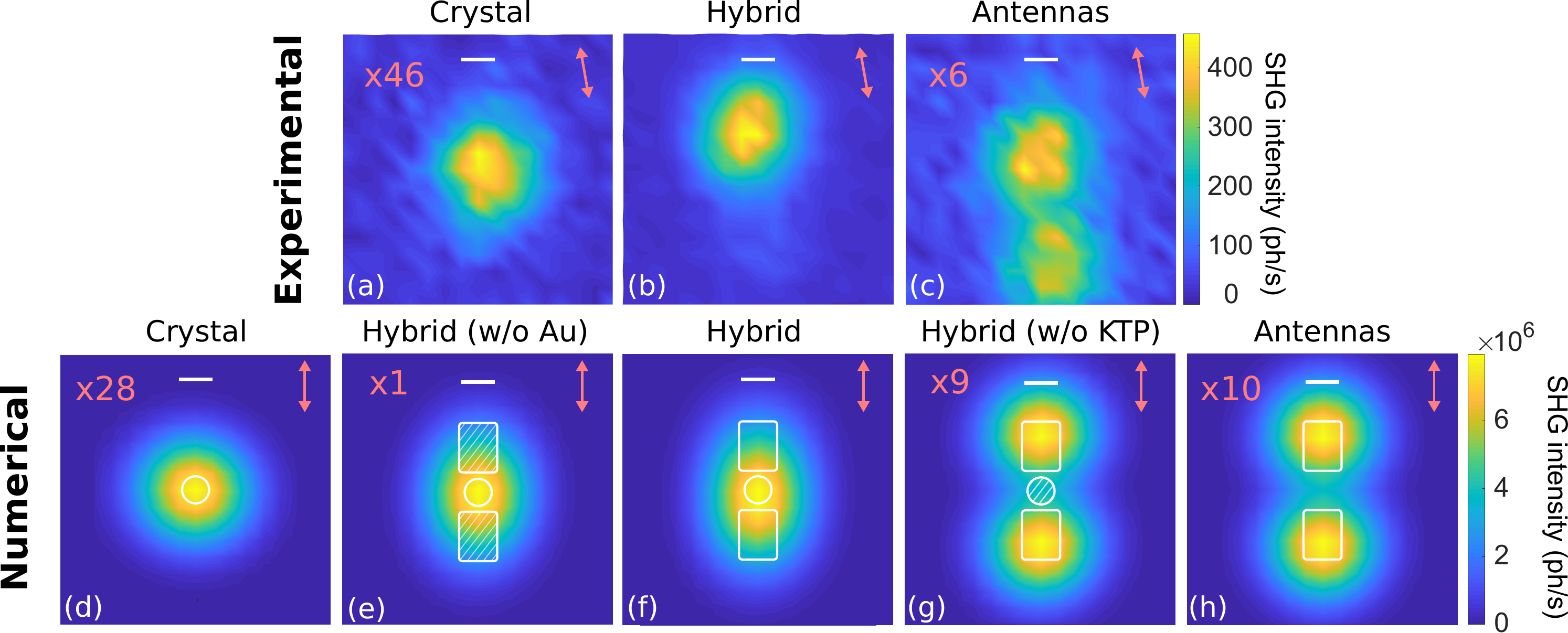}
  \caption{(a-c) Normalized experimental and (d-h) numerical SHG excitation cartographies of (a,d) a bare KTP crystal, (c,h) bare gold antennas and (b,f) hybrid Au-KTP structure. Parameters in simulation are from SEM images. (e) Hybrid SHG intensity without the nonlinear response of antennas, and (g) without the one from the crystal. Colorbars are the same for experimental  and numerical data, respectively, with each normalization factors indicated. Scale bar: 100 nm.}
  \label{fig:comparison}
\end{figure}

Let us now focus on the question of the origin of the measured enhancement. Two (non-exclusive) hypotheses can be made: either (\textbf{H1}) the nonlinear response of the antennas themselves is increased because of the dielectric environment change induced by the crystal placed in the antenna gap, or (\textbf{H2}) the response of the crystal itself is enhanced by the plasmon resonances of the antennas (linear response). So far, both hypotheses have received supporting data from the literature, with Linnenbank et al \cite{Linnenbank2016} explaining their observed 2-fold enhancement by (\textbf{H1}), Zhou et al \cite{Zhou2010} and Utikal et al \cite{Utikal2011}  by (\textbf{H2}) whereas Hentschel et al \cite{Hentschel2016} conclude that both effects are present simultaneously and cannot be ignored from each other. As a consequence, careful comparison has to be made with respect to simulations where contributions from both the nonlinear crystal and the plasmonic antennas can be taken into account simultaneously or separately. To this end, we compare in Figure \ref{fig:comparison} the experimental SHG response on excitation cartography for (a) the individual crystal (before hybrid structure fabrication), (b) the hybrid structure with the same crystal, and (c) Au antennas with the same drawing pattern fabricated nearby without crystal. The red arrow indicates the linear polarization of the incident light. As indicated, a 46-fold enhancement is observed for the hybrid structure compared to the crystal, while antennas have a 6 times lower SHG intensity than the hybrid. These numbers, which are in the range between the reported 15-fold enhancement from colloidal single hybrid dimers \cite{Timpu2017} and the 500-fold increase in core-shell Au-BaTiO$_3$ particles \cite{Pu2010}, are covered by the scatter plot in graphs of Figure \ref{fig:comparison}(b) and (c), thus showing good agreement with previous results in the literature for similar systems.

In order to identify the origin of the nonlinear efficiency increase and rule out one hypothesis or the other, we have run simulations corresponding to each of the experimental cases, respectively represented on Figures \ref{fig:comparison}(d), (f) and (h) for gold-KTP structures. Predicted intensities, expressed with respect to simulated hybrid structure intensity, show the correct orders of magnitude in enhancements (see the multiplication factors in each panel). The cartography maps are also in fair agreement. The slight differences can also be interpreted as the result of shape differences between the real structure and the simulated one (size, form and gap), as evidenced by the large influence of these parameters as shown in Figure \ref{fig:comparison}(a). Further information is obtained by simulating the hybrid structure without nonlinear response either from the antennas or from the crystal, while still keeping the linear properties of each material. Figures \ref{fig:comparison}(e) and (g) show the corresponding maps: removing the nonlinearity of the plasmonic antennas does not change the overall SHG intensity nor the cartography shape, whereas keeping only linear dielectric properties for the crystal reduces the SHG intensity by a factor of 9 and changes dramatically the SHG excitation map from 1 lobe to 2 lobes. In accordance with the experimental data, these results show undoubtfully that the SHG enhancement observed in single hybrid structures made of KTP nanocrystal and Au dimer is attributed to the presence of resonant antennas near the nonlinear crystal according to hypothesis (\textbf{H2}).

The case of aluminum based hybrid structures is drastically different. Indeed, while bulk contribution $d$ in equation \ref{eq:bulk} is of the order of 1 in gold, the value of the surface contribution $a$ of equation \ref{eq:surf} in aluminum is -22 according to experimental data by Murphy \textit{et al} \cite{Murphy1989}. This effect, combined with double resonant conditions achievable in aluminum antennas\cite{EthisDeCorny2016}, explains why simulations predict a stronger nonlinear contribution from plasmonic antennas. As a result, the same analysis as in Figure \ref{fig:comparison} enables us to conclude that in an optimized aluminum-KTP structure, the main contribution to the SHG comes from the plasmonic antennas themselves, in accordance with hypothesis \textbf{H1}, since removing the nonlinearity from the KTP crystal does not significantly change the total SHG intensity. This also means that the enhancement factor between bare antennas and hybrid structure is of the order of 1, thus limiting the interest of hybrid structure design in terms of efficiency magnification for the case of aluminum. Still, aluminum-based design shows higher nonlinear conversion efficiency under our experimental conditions as compared with gold-based structures.

%[IL VAUT MIEUX AVOIR LA NOUVELLE FIGURE 4 POUR DISCUTER] expliquer que a = -22 fait que la réponse de l'antenne est comparable à celle de l'hybride (i.e. crystal génère peu) Donc au final H1 pour Al mais mitiger en disant que exaltation moins forte que l'or mais signal absolu plus grand.

%Crystals, experiment characteristics (wavelength, power, NA, pulsed, polarization), KTP / characteristics / synthesis, method for 3D orientation characterization.

%Hybrid structures design, from simulations, plasmonic structures along main (z) axis in sample plan, reference antennas nearby.

%Parameters of plasmonic structures determined after numerical optimisation, see Figure for kinds of structures obtained (including dilated ones), gold vs aluminum, general results: only >1 for both crystals and antennas (one point under: not sure crystal still there), gold less self NL but good enhancement, better enhancement with dilated Al, in agreement with simulations.

%Pinpoint origin: compare KTP, antennas, and hybrids. Enhancement vs each... not enough. -> Simulations: turn on or off NL contributions, from antennas or crystal, and see what happens, result: no antenna=same, no crystal=:10, antennas play by their linear field enhancement and far-field coupling (?? gold), not by their NL! => origin=NL crystal + L antennas, for gold at least...?

\section{Conclusion}

We have demonstrated that plasmonic hybrid structures coupling a plasmonic antenna and a SHG emitting nanoparticle can lead to an enhancement of the SHG emission with respect to the one of each component. On the one hand, for gold dimer and KTP crystal combination, comparison with simulations attribute unequivocally the origin of this enhancement to the plasmonic resonance in the antenna, which, when placed in the vicinity of a nonlinear crystal, increases the intrinsic response of the latter and helps it couple to the far-field. On the other hand, the same comparison between experiment and numerical simulations for aluminum-KTP structures show that the nonlinear contribution of the plasmonic antennas is superior to the enhancement effect on the crystal in this case, due to both stronger intrinsic nonlinear susceptibility of aluminum and double resonant conditions in nanodimers. These results rely on the optimization of each parameter of the structure thanks to far-field crystallographic analysis and FEM simulation tools that successfully predict the nonlinear response of each component quantitatively, without any free parameter. This work introduces a general framework that can be used for different geometries, materials or phenomena. One particular finding is that the closer the crystal from the antenna(s), the larger the enhancement factor. Therefore, configurations such as patch antennas where the crystal is on top or under a plasmonic antenna, could be particularly interesting to study. In addition, more systematic fabrication methods developped in recent years can turn the fabrication protocol presented here into a systematic approach, paving the way to rigorous study of different parameters on the performances of hybrid structures. Finally, the obtained results can be transferred to other nonlinear optical phenomena such as non-degenerate Sum Frequency Generation, or the reverse phenomenon, namely Spontaneous Parametric Down Conversion which has not yet been observed in hybrid structures.

%Unequivocally shown that hybrids do better than each, 1+1>2, comes from L antenna resonance + NL crystal, also configuration, parameters tuning
%For the next: search for appropriate shape, change material (Al with good parameters), and above all, switch to SPDC, since plasmonic antennas are already ready for this anyway....

\section{Experimental section}

%Summary of setup and fabrication methods... as well as simulations?

%Experiments: We have used a Ti-Sa femtosecond pulsed laser source (MaiTai from Spectra-Physics), with repetition rate of 80 MHz, pulse width of about 300 fs at the sample level, working at 850 nm central wavelength and average power of 100 $\mu$W at the sample level. To make observations at the single particle level, we need strong focusing (objective, NA...)

%laser parameters, optical configuration, spectrometer, acquisition, polar
%Fabrication: e-beam lithography, KTP synthesis, substrate, development?
%Simulations: contributions prises en compte, principe de fonctionnement?

In order to reach the single particle level in nonlinear optical studies, the experimental setup needs to reach diffraction-limited spatial resolution, nanometer positioning precision, high light sensitivity and low optical noise level. Each part of the setup used in this study has been designed to meet these goals. The laser source is a Ti:Sapphire working on a 100-femtosecond pulsed mode with a repetition rate of 80 MHz and a central wavelength of 850 nm. The average optical power for SHG experiments is set at 100 $\mu$W by using variable neutral density filters in order to avoid optical damage for plasmonic structures. We use an immersion oil microscope objective (x100, 1.3 numerical aperture) for the focusing of the laser beam as well as the collection of signals in reflexion mode, providing diffraction-limited spot size of 330 and 165 nm respectively at the excitation (fundamental) and harmonic frequencies. SHG photons are separated from the input beam optical axis by using 685 nm single-edge dichroic beamsplitter (DeltaMicroscopy FF685-Di02). To compensate for chromatic aberrations from the microscope objective, remote-controlled telescopes are used so as to make both excitation and signal wavelength focal planes overlapping at the surface of the sample. Regarding nano-positioning of the sample, closed-loop piezoelectric translation stages are set under it to provide lateral scanning capabilities with nanometer precision with respect to the laser beam focal spot. Coupled with a tracking algorithm, the full computer control of the setup automatically optimizes the optical signal by tuning the positioning and chromatic aberration compensation before each set of measurements, thus providing optimized and reproducible experimental conditions. The collected light is then spectrally selected by a monochromator with 150 g/mm grating and analyzed through a polarizing beam-splitter coupled to two low-noise Avalanche Photodiode modules (PerkinElmer SPCM-AQR-16). An ultra-fast acquisition card (Fast ComTec MCS6A) is synchronized with the laser pulses to discriminate between time-correlated SHG signals and ambient or dark-noise, reducing average dark count rates to a few Hz for each module.

KTP nanocrystals were synthesized using a procedure previously reported in \cite{Mayer2013}. Briefly, titanium butoxide was mixed with HCl in distilled water. A solution of KH$_2$PO$_4$ in distilled water was then added. After pH adjustment at 6 and stirring for 24 hours, the precipitate was centrifuged and dried at 100$^{\circ}$C. It was then annealed at 700$^{\circ}$C for 2 hours, allowing the crystallization of the KTiOPO$_4$ (KTP) phase. Thanks to the presence of excess KCl, further removed by dissolution in water, the crystallization is limited, leading to nanosized KTP crystals.

The sample used in our experiments consists in a 170 $\mu$m-thick glass substrate (Zeiss 1.5 coverslip). The plasmonic structures are fabricated using standard electron beam lithography, using exactly the same protocol for aluminum and gold structures. After cleaning with acetone in an ultrasound bath and rinsed with isopropanol and deionized water, a 3-minutes oxygen plasma cleaning removes any organic residue on the surface. After this, a 3\% PMMA layer is deposited and spin-coated on the surface before 180$^{\circ}$C baking for 5 minutes. A 15-nm thick gold layer is deposited on the resist by electron-gun evaporation to evacuate charges during electronic lithography, because of the dielectric nature of the substrate. After lithography, the gold layer is chemically removed using KI/I$_2$ solution. The resist is then developed in a 0$^{\circ}$C isopropanol-deionized water (3:1) solution before isopropanol rinsing. Finally, a 35-nm thick metallic layer (gold or aluminum) is deposited onto the sample using electron-gun evaporation, before N-methyl-2-pyrrolidone lift-off at 80$^{\circ}$C. All measurements performed on nanocrystals and plasmonic structures have been realized using scanning electron microscopy (SEM), with a typical precision of $\pm$3 nm on the values (determined by successive measurement performed on the same structures), while thickness has been determined using atomic force microscopy.

Once the realignment gold structures have been realized on the glass substrate by electron beam lithography, deposition of the crystals is made by spin-coating an aqueous solution containing the nanocrystals onto the substrate at room temperature and at a speed of 6000 rpm for 30 seconds, after oxygen plasma exposition of the substrate to make it more hydrophilic. After this, SEM imaging enables us to localize areas of interest containing enough crystals to get many potential hybrid structures, while not too densely to avoid several-crystals agglomerates that are not useful for our purpose. In total, 122 crystals have been observed in 4 areas and studied using SHG polarization analysis, from which 122 hybrid structures have been realized, including 80 exploitable ones. The other ones either had their crystal "disappear" on SEM images after fabrication (probably being under the plasmonic antennas), or were in an area where the lift-off step failed.

Numerical simulations are based on a four step procedure. First, the local excitation fields at the fundamental frequency are evaluated using analytic methods accounting for the tight focusing by the microscope objective, the presence of immersion oil as well as the refraction induced by the glass substrate \cite{Novotny2012}. Then, Maxwell's equations are numerically solved at the fundamental frequency within the scattered field formulation to determine the electric field in the vicinity of the nanostructure. The linear optical properties of the latter are described using tensorial dielectric functions, in agreement with the crystal anisotropy. In the third step, the nonlinear source current are computed using either the Rudnick and Stern parameters \cite{Rudnick1971, Teplin2002, Bachelier2010, Kauranen2012} for the plasmonic antennas or the nonlinear susceptibility tensor for the nonlinear crystal\cite{Pack2004}. This allows solving Mawxell's equations at the harmonic frequency using here a soft formulation to properly include the nonlinear currents, avoiding thereby continuity issues for surface contributions. Finally, the fourth step consists in analytically propagating the harmonic electric field back through the substrate and collection microscope objective (using a Green's tensor framework) and focusing it onto the detector\cite{Novotny2012} in order to quantitatively evaluate the measured SHG signal. Note that there is no adjustable parameter in these simulations as the dielectric functions and nonlinear susceptibilities are taken from tabulated values and the antenna parameters are deduced from AFM (height) or SEM (length and width) images. The experimental transmission through the entire optical setup as well as the detector efficiency are not included in order to provide a general validity to the presented simulations.

\begin{acknowledgement}

The authors acknowledge the financial support of the Agence Nationale de la Recherche (Grant ANR-14-CE26-0001-01-TWIN), the Chaire IUA awarded to Guillaume Bachelier by the Université Grenoble Alpes and the Ph.D. grant to Nicolas Chauvet from the Laboratoire d'excellence LANEF (ANR-10-LABX-51-01). The authors thank the Nanofab team at Institut Neel for the nanofabrication platform and their strong support.

\end{acknowledgement}

%%%%%%%%%%%%%%%%%%%%%%%%%%%%%%%%%%%%%%%%%%%%%%%%%%%%%%%%%%%%%%%%%%%%%
%% The same is true for Supporting Information, which should use the
%% suppinfo environment.
%%%%%%%%%%%%%%%%%%%%%%%%%%%%%%%%%%%%%%%%%%%%%%%%%%%%%%%%%%%%%%%%%%%%%
%\begin{suppinfo}
%
%A listing of the contents of each file supplied as Supporting Information
%should be included. For instructions on what should be included in the
%Supporting Information as well as how to prepare this material for
%publications, refer to the journal's Instructions for Authors.
%
%The following files are available free of charge.
%\begin{itemize}
%  \item Filename: brief description
%  \item Filename: brief description
%\end{itemize}
%
%\end{suppinfo}

%%%%%%%%%%%%%%%%%%%%%%%%%%%%%%%%%%%%%%%%%%%%%%%%%%%%%%%%%%%%%%%%%%%%%
%% The appropriate \bibliography command should be placed here.
%% Notice that the class file automatically sets \bibliographystyle
%% and also names the section correctly.
%%%%%%%%%%%%%%%%%%%%%%%%%%%%%%%%%%%%%%%%%%%%%%%%%%%%%%%%%%%%%%%%%%%%%
%\bibliographystyle{unsrtnat}%!! fichier .bst modifié !!
\bibliography{./article-hybrides}

\end{document}